\documentclass[final]{aipproc}
\layoutstyle{8x11single}
\newcommand{\be}{\begin{eqnarray}}
\newcommand{\ee}{\end{eqnarray}}
\newcommand{\non}{\nonumber \\}

\begin{document}

\title{Meson-baryon interaction in the meson exchange picture}


\author{M.~D\"oring}{
  address={Institut f\"ur Kernphysik and J\"ulich Center for Hadron Physics, \\
Forschungszentrum J\"ulich, D-52425 J\"ulich,Germany}
}

%\author{C.~Hanhart}{
%  address={Institut f\"ur Kernphysik and J\"ulich Center for Hadron Physics, \\
%Forschungszentrum J\"ulich, D-52425 J\"ulich,Germany}
%}

%\author{F.~Huang}{
%  address={Department of Physics and Astronomy, University of Georgia, Athens, Georgia 30602, USA}
%}

%\author{S.~Krewald}{
%  address={Institut f\"ur Kernphysik and J\"ulich Center for Hadron Physics, \\ 
%Forschungszentrum J\"ulich, D-52425 J\"ulich,Germany}
%}

%\author{U.-G.~Mei\ss ner}{
%  address={Helmholtz-Institut f\"ur Strahlen- und Kernphysik (Theorie) and Bethe Center for Theoretical
%Physics, \\
%Universit\"at Bonn, Nu\ss allee 14-16, D-53115 Bonn, Germany}
%,altaddress={Institut f\"ur Kernphysik and J\"ulich Center for Hadron Physics, \\ 
%Forschungszentrum J\"ulich, D-52425 J\"ulich,Germany}
%}

%\author{D.~R\"onchen}{
%  address={Institut f\"ur Kernphysik and J\"ulich Center for Hadron Physics, \\ 
%Forschungszentrum J\"ulich, D-52425 J\"ulich,Germany}
%}

\begin{abstract}
Elastic $\pi N$ scattering and the reaction $\pi^+ p \to K^+\Sigma^+$ are 
described simultaneously in a
unitary coupled-channels approach which respects analyticity. SU(3) flavor symmetry is used to relate the
$t$- and $u$- channel exchanges that drive the meson-baryon interaction in the different channels. Angular
distributions, polarizations, and  spin-rotation parameters are compared with available experimental data.
The pole structure of the amplitudes is extracted from the analytic continuation. 
\end{abstract}

\maketitle

%%%%%%%%%%%%%%%%%%%%%%%%%%%%%%%%%%%%%%%%%%%%
%% MAINMATTER
%%%%%%%%%%%%%%%%%%%%%%%%%%%%%%%%%%%%%%%%%%%%

\section{Introduction}
The excitation spectrum of baryons and mesons is expected to reveal important information on the mechanism
of confinement as well as the instrinsic structure of hadrons. Properties of baryon resonances have been
obtained by lattice calculations ~\cite{Durr:2008zz,Bulava:2010yg,Engel:2010my}, mostly for the
ground states but also for some excited states~\cite{Bulava:2010yg}.  In quark
models~\cite{Isgur:1978xj,Capstick:1986bm,Loring:2001kx},  a rich spectrum of excited states is predicted.
Many of these resonances    could be identified in elastic $\pi N$ scattering,  while at higher energies,
usually more states are predicted than seen, a fact commonly referred to as the ``missing resonance
problem''.  Since resonances not seen in the $\pi N$ channel might predominantly couple to other channels,
there are intensive experimental efforts to measure, among others, multi-pion or $KY$ final states, where
$KY=K\Lambda$ or $K\Sigma$.

 The reaction $ \pi^+  p \rightarrow K^+ \Sigma^+$ provides access to a pure isospin $I=3/2$ two-body
reaction channel in meson-nucleon dynamics. Moreover, the weak decay $\Sigma^+ \rightarrow p \pi^0$ 
allows to determine the polarization of the produced $ \Sigma^+ $.
Thus, we have performed a first combined analysis of the reactions $\pi N\to\pi N$ and $\pi^+ p\to K^+\Sigma^+$ within the
unitary dynamical coupled-channels framework~\cite{Doring:2010ap}, with the results summarized in these proceedings.
 
Dynamical coupled-channels models~\cite{Schutz:1998jx,Krehl:1999km,Gasparyan:2003fp,Doring:2009yv,JuliaDiaz:2007kz,Suzuki:2009nj,Tiator:2010rp} 
are particularly suited for combined data analyses: E.g., in the work presented here~\cite{Doring:2010ap}, 
based on the ``J\"ulich model''~\cite{Schutz:1998jx,Krehl:1999km,Gasparyan:2003fp,Doring:2009yv},
the SU(3) fla\-vor symmetry for the exchange processes allows to relate different final states. The $t$- and $u$-channel diagrams
connect also different partial waves and the respective backgrounds. 
Thus, the treatment of the interaction via meson and baryon exchange is expected to lead to a
realistic background, with strong restrictions on the free parameters.
In view of this, the strategy to perform baryon spectroscopy is to introduce only a minimum number of bare
resonance states in order to obtain a good description of the data.

The coupled-channels scattering
equation~\cite{Schutz:1998jx,Krehl:1999km,Gasparyan:2003fp,Doring:2009yv} used in the present
formalism fulfills  two-body unitarity, as well as some requirements of three-body unitarity.  
Furthermore, it fulfills analyticity and takes into account the dispersive parts
of the intermediate states as well as the off-shell behavior dictated by the interaction  Lagrangians. This
integral equation which is solved in the $JLS$-basis is given by
\be
&&\langle L'S'k'|T_{\mu\nu}^{IJ}|LSk\rangle=\langle L'S'k'|V_{\mu\nu}^{IJ}|LSk\rangle\non
&&+\sum_{\gamma\, L''\, S''}\int\limits_0^\infty k''^2\,dk''
\langle L'S'k'|V_{\mu\gamma}^{IJ}|L''S''k''\rangle
\,\frac{1}{z-E_{\gamma}(k'')+i\epsilon}\,\langle L''S''k''|T_{\gamma\nu}^{IJ}|LSk\rangle
\label{bse}
\ee
where $z$ is the scattering energy,  $J\,(L)$ is the total angular (orbital angular) momentum, $S\,(I)$ is the total spin (isospin),
$k(k',\,k'')$ are the incoming (outgoing, intermediate) momenta,  and $\mu,\,\nu,\,\gamma$ are channel
indices. In Eq.~(\ref{bse}),
$E_\gamma$ is the on-mass shell energy in channel $\gamma$~\cite{Doring:2009yv}. 
The pseudo-potential $V$ iterated in Eq. (\ref{bse}) is constructed from an
effective interaction based on the Lagrangians of Wess and Zumino \cite{Wess:1967jq,Meissner:1987ge},
supplemented by additional terms \cite{Krehl:1999km,Gasparyan:2003fp} for including the $\Delta$ isobar, the
$\omega$, $\eta$, $a_0$ meson, and the $\sigma$. The channel space is given by $N \pi, N\eta,
N\sigma, \Delta \pi$, $N \rho$~\cite{Schutz:1998jx,Krehl:1999km,Gasparyan:2003fp} and --- 
as the novelty presented here --- $\Lambda K$ and $\Sigma K$~\cite{Doring:2010ap}.

\begin{figure}
%\begin{center}
\includegraphics[height=0.2\textwidth]{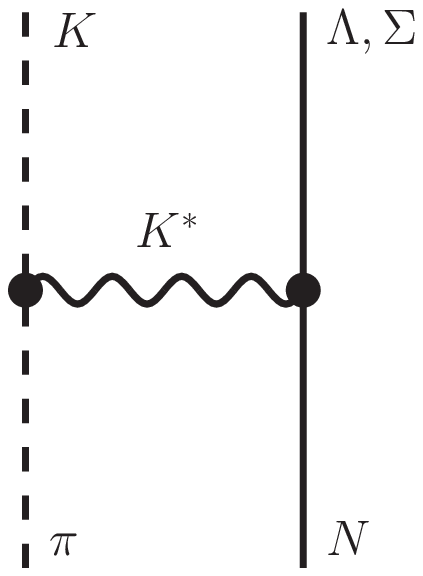} %\hspace*{-0.24cm}
\includegraphics[height=0.2\textwidth]{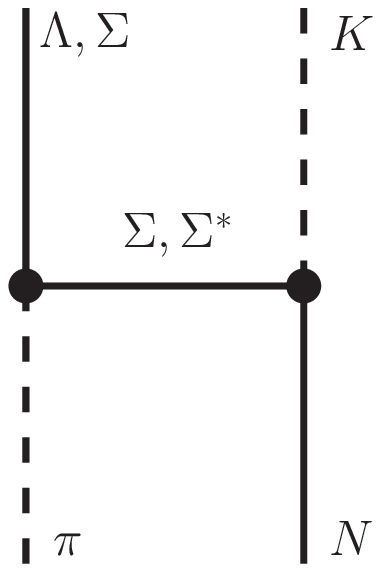}  %\hspace*{-0.24cm}
\includegraphics[height=0.2\textwidth]{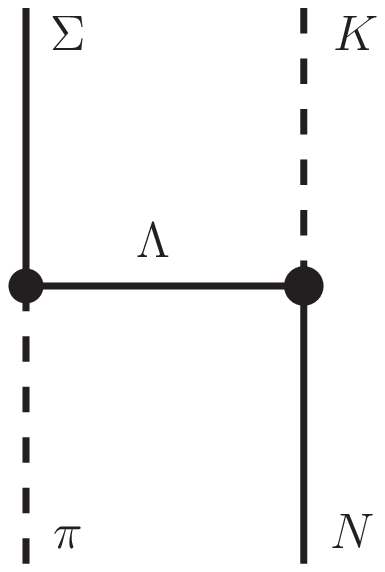} \\
\includegraphics[height=0.2\textwidth]{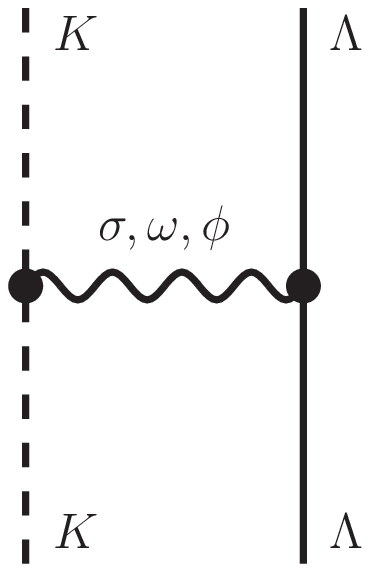}   %\hspace*{-0.24cm}
\includegraphics[height=0.2\textwidth]{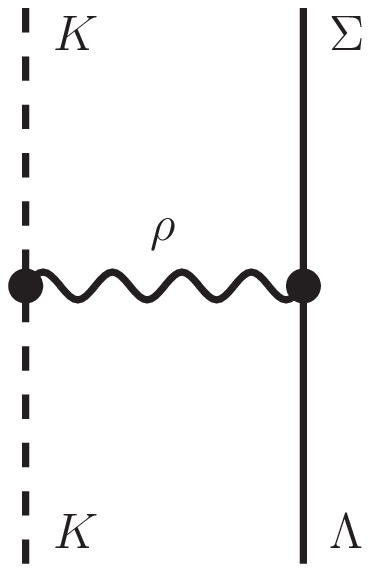}  %\hspace*{-0.24cm}
\includegraphics[height=0.2\textwidth]{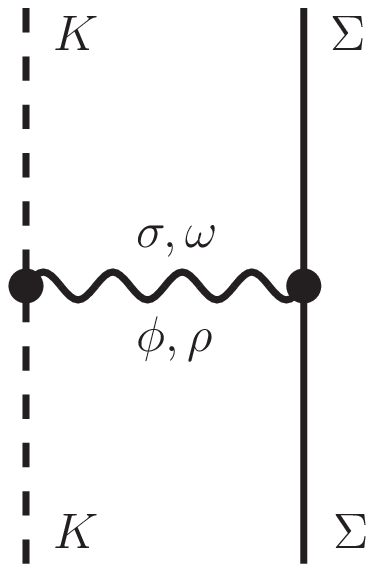}
%\end{center}
\caption{$\pi N \to KY$ and $KY\to KY$ transitions. For the other transitions used in this approach, 
see Refs.~\cite{Krehl:1999km,Gasparyan:2003fp}.}
\label{fig:kydiagrams}     
\end{figure}

The $t$- and $u$-channel processes provide the non-resonant interaction in the meson exchange picture. 
The transition potentials without participation of $KY$ have been derived in
Refs.~\cite{Krehl:1999km,Gasparyan:2003fp} and explicit expressions can be found in these
references. In Fig.~\ref{fig:kydiagrams}, we show the extension to the $KY$ channels.
Most of the vertices present in these diagrams are related to the ones of Refs.~\cite{Krehl:1999km,Gasparyan:2003fp}
using SU(3) symmetry; details can be found in Ref.~\cite{Doring:2010ap}. 

Apart from this non-resonant interaction, $s$-channel diagrams have been introduced, which can be regarded as resonances 
( for details, see Ref.~\cite{Doring:2010ap}. In the present study, we concentrate on the reaction $\pi^+p\to K^+\Sigma^+$
and the isospin $I=3/2$ elastic $\pi N$ scattering. Thus, 
one bare $s$-channel state is included in each of the $I=3/2$ partial waves 
S31, P31, D33, D35, F35, F37. Two are present in P33. 
These states were allowed to couple to all $I=3/2$ channels 
$\pi N$, $K\Sigma$, $\pi\Delta$ and $\rho N$. Together with these four bare couplings, the bare mass has to be 
left free as a fit parameter. Thus, there are altogether 40 parameters, apart from some free parameters in the $t-$ and $u-$channel
diagrams; the latter parameters in the new diagrams from Fig.~\ref{fig:kydiagrams} have been chosen closely to their counterparts
without $KY$ channels from Refs.~\cite{Gasparyan:2003fp}. Thus, in the present approach there is explicit SU(3) breaking from the different hadron 
masses, but also a small SU(3) breaking from the $t$- and $u-$channel diagrams, mostly in terms of different form factors~\cite{Doring:2010ap}.

%%%%%%%%%%%%%%%%%%%%%%%%%%%%%%%%%%%%%%%%%%%%%%%%%%%%%%%%%%%%%%%%%%%%%%%%%%%%%%%%%%%%

\section{Results}
\begin{figure}
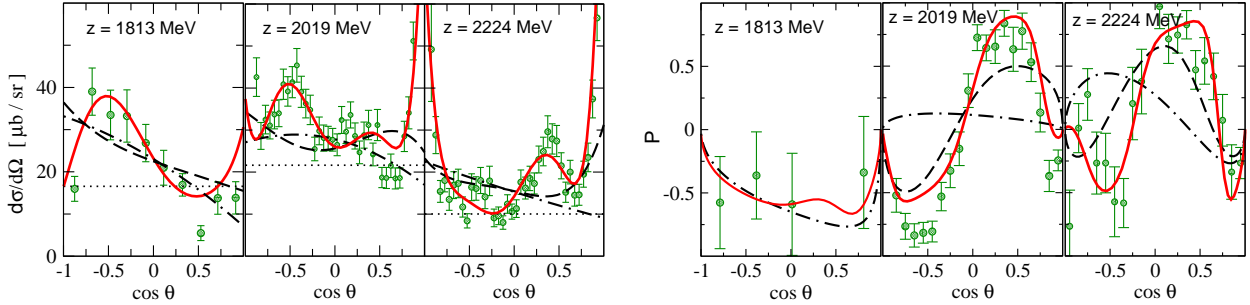

%\begin{center}
\includegraphics[height=0.242\textwidth]{graphs/dsdo_k+s+_low_spd.eps}  \hspace*{-0.24cm}
\includegraphics[height=0.242\textwidth]{graphs/dsdo_k+s+_up_spd.eps}   \hspace*{-0.24cm} 
\includegraphics[height=0.242\textwidth]{graphs/dsdo_k+s+_high_spd.eps} \hspace*{0.3cm}
\includegraphics[height=0.242\textwidth]{graphs/pola_k+s+_low_spd.eps}  \hspace*{-0.24cm}
\includegraphics[height=0.242\textwidth]{graphs/pola_k+s+_up_spd.eps}   \hspace*{-0.24cm}
\includegraphics[height=0.242\textwidth]{graphs/pola_k+s+_high_spd.eps} 
%\end{center}
\caption{Selected results of the presented study (red lines) for three
typical energies, for differential cross section and polarization. Data: Ref.~\cite{Candlin:1982yv} and references therein. 
Also, contributions from different partial waves are shown: Dotted lines: $S$ wave. Dash-dotted lines: $S+P$ waves. Dashed lines: $S+P+D$ waves.}
\label{fig:spd}     
\end{figure}

Results for three typical scattering energies $z$ are shown in Fig.~\ref{fig:spd}. The full results for differential cross section $d\sigma/d\Omega$ and polarization $P$, 
at 32 different energies from the $K\Sigma$ threshold up to $z=2.3$ GeV, can be found in Ref.~\cite{Doring:2010ap}.
The red solid lines show the result of this study. Overall, the data are well described over the entire
energy range. For energies above 2 GeV, we do not claim
validity of the present model, because the analysis of Refs.~\cite{Gasparyan:2003fp, Doring:2009yv} has been
limited to that energy. Consequently, at the highest energies, the $K^+\Sigma^+$ data have not been fitted, 
but up to $z\sim 2.25$ GeV the description of the data is 
still good, as Fig.~\ref{fig:spd} shows. 
In Fig.~\ref{fig:spd}, also the influence of individual partial waves is illustrated. 
At lower energies, $S-P$ wave interference is enough to describe the polarization, but not entirely the 
differential cross section (dash-dotted lines at $z=1813$ MeV). At higher energies, the figure shows that 
indeed all partial waves are needed to quantitatively describe the data; in particular, F waves are important.

\begin{figure}
%\begin{center}
\includegraphics[width=0.53\textwidth]{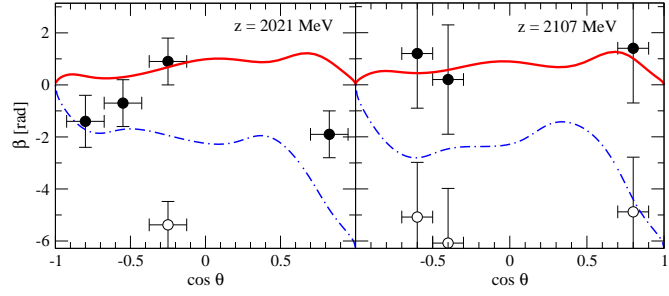} 
%\end{center}
\caption{Spin rotation parameter $\beta$ of $\pi^+ p \rightarrow K^+\Sigma^+$ at $z=2021$ and $z=2107$ MeV.
Note that $\beta$ is $2\pi$ cyclic which leads to additional data points at shifted values shown by the
empty circles. Data: Ref.~\cite{Candlin:1988pn}. (Red) solid lines: Prediction from present solution. 
(Blue) dash-dotted lines: Prediction from Ref.~\cite{Candlin:1983cw}.}
\label{fig:spinrot}     
\end{figure}

\begin{figure}
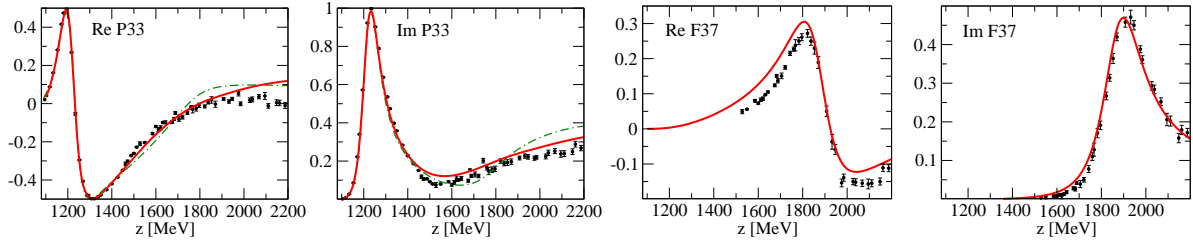
 
%\begin{center}
\includegraphics[width=0.48\textwidth]{graphs/p33.eps} 
\includegraphics[width=0.47\textwidth]{graphs/f37.eps}
%\end{center}
\caption{Elastic $\pi N\to\pi N$ partial waves P33 and F37. (Red) solid lines: 
Present solution. (Green) dash-dotted lines: J\"ulich model, 
solution 2002 from Ref.~\cite{Gasparyan:2003fp}. For the other partial waves see Ref.~\cite{Doring:2010ap}. 
Data points: GWU/SAID~\cite{Arndt:2006bf}.}
\label{fig:pw1}       
\end{figure}

Fig.~\ref{fig:spinrot}  shows the predicted spin-rotation parameter $\beta$~\cite{Doring:2010ap} (solid lines). 
The prediction from the isobar analysis of  Ref.~\cite{Candlin:1983cw} is also
shown. 
It is interesting to note that while $d\sigma/d\Omega$ and $P$ are described to a similar precison in both studies,
the outcome for $\beta$ is so different. This demonstrates that higher precision data for $\beta$ would help further pin down 
the partial wave content. In any case, what distinguishes the present analysis~\cite{Doring:2010ap} from 
Ref.~\cite{Candlin:1983cw} is (apart from conceptual differences) the simultaneous consideration of  
elastic $\pi N$ scattering. The results of the presented study~\cite{Doring:2010ap} for the P33 and F37 
partial waves are shown in Fig.~\ref{fig:pw1}, for the other partial waves see Ref.~\cite{Doring:2010ap}.
The partial waves for the reaction $\pi^+p\to K^+\Sigma^+$ are shown in Ref.~\cite{Doring:2010ap}.

The resonance content of the amplitude is extracted using the analytic continuation following 
Refs.~\cite{Doring:2009yv,Doring:2009bi} (see also Refs.~\cite{Doring:2009uc,Doring:2010rd}). In Table~\ref{tab:bra2}, pole positions, residues, and branching ratios
for the most prominent resonances in the reaction $\pi^+p\to K^+\Sigma^+$ are shown and compared to results from Refs.~\cite{Candlin:1983cw,Penner:2002ma,Shklyar:2004dy}. 
\begin{table}
\caption{Left: Resonance pole positions $z_0$ in the presented study~\cite{Doring:2010ap}.
Center: $\pi^+ p\to K^+\Sigma^+$ residues from Ref.~\cite{Doring:2010ap}. Right: 
Transition branching ratios.}
%\begin{center}
\renewcommand{\arraystretch}{1.15}
   \begin {tabular}{cc|c|ccc} %\hline\hline
%\multicolumn{5}{c}{$\pi^+ p\to K^+\Sigma^+$}\\  \hline
			& Re $z_0$~[MeV]	&$|r|$ [MeV]		&\multicolumn{3}{c}{$(\Gamma^{1/2}_{\pi N}\Gamma^{1/2}_{K\Sigma})/\Gamma_{\rm tot}$ [\%]}\\ 
			& -2 Im $z_0$~[MeV]	&$\theta$~[$^0$] 	&Ref.~\cite{Doring:2010ap}&Ref.~\cite{Candlin:1983cw}&Refs.~\cite{Penner:2002ma,Shklyar:2004dy}\\
\hline
$ \Delta(1905)F_{35}$  &1764	&1.4    & 1.23  & 1.5(3)	& $<$1  \\
5/2$^+$ ****           & 218	& -313  &       &	&   	\\
\hline
$ \Delta(1910)P_{31}$  &1721	& 5.5   & 2.98  & $<$3	& 1.1   \\
1/2$^+$ ****           &323 	& -6    &       &	&   	\\
\hline
$ \Delta(1920)P_{33}$  &1884	& 5.9   & 5.07  & 5.2(2)	& 2.1(3)\\
3/2$^+$ ***            & 229	& -38   &       &	&   	\\
\hline
$ \Delta(1930)D_{35}$  &1865	& 1.6   & 2.14  & $<$1.5	&       \\
5/2$^-$ ***            & 147	& -43   &       &	&	\\
\hline
$ \Delta(1950)F_{37}$  &1873	& 2.7   & 2.54  & 5.3(5)	& ---   \\
7/2$^+$ ****           & 206	& -255  &       &	&  	\\
\hline
%\hline
\end {tabular}
%\end{center}
\label{tab:bra2}
\end{table}
The full list of resonances and their properties can be found in Ref.~\cite{Doring:2010ap}. As for the well-established 4-star resonances, 
it is no surprise that most of the pole positions found in this study
are in good agreement with the values from the GWU/SAID analysis~\cite{Arndt:2006bf}, because the
partial waves from that analysis serve as input for the present study. However, the $\Delta(1920)P_{33}$ and $\Delta(1930)D_{35}$ 
resonances show no or very small resonance signals in the
GWU/SAID analysis of elastic $\pi N$ scattering. Their position is, thus, barely
fixed from elastic $\pi N$ scattering.  It is then interesting to note that the constraints from the
$K^+\Sigma^+$ data  lead to resonance positions in vicinity to those quoted in the PDG~\cite{pdg}, rated with 3 stars.
Thus, we can accumulate further evidence for these states and their positions.
Also, we find a $\Delta(1600)P_{33}$ resonance, 
dynamically generated mainly from the $\pi\Delta$ channel (see also Ref.~\cite{Oset:2009vf}). Qualitatively, it shares
properties with the corresponding state quoted in the PDG~\cite{pdg} like a strong decay into the $\pi\Delta$ channel.

In summary, a first combined analysis of the reactions $\pi N\to\pi N$ and $\pi^+ p\to K^+\Sigma^+$ within the 
dynamical coupled-channels framework has been presented. In this approach,
 for both $\pi N$ and $K\Sigma$ a realistic and structured background can be provided. 
Consequently, only a minimal set of $s$-channel resonances
is needed to obtain a good fit to the combined data sets.
This is also tied to the fact that in this field-theoretical, Lagrangian based approach, the dispersive
parts are fully included and thus, analyticity is ensured.
Apart from the well-established 4-star resonances, there is a
clear need for the three-star $\Delta(1920)P_{33}$ resonance. This state is found to couple only weakly to $\pi N$ but
stronger to $K\Sigma$, and thus evidence for a ``missing resonance state'' could be
accumulated.

%%%%%%%%%%%%%%%%%%%%%%%%%%%%%%%%%%%%%%%%%%%%%%%%
%% BACKMATTER
%%%%%%%%%%%%%%%%%%%%%%%%%%%%%%%%%%%%%%%%%%%%%%%%

\begin{theacknowledgments}
The work of M.D. is supported by DFG (Deutsche Forschungsgemeinschaft, GZ: DO 1302/1-2). This work is
supported in part by the Helmholtz Association through funds provided to the virtual institute ``Spin and
Strong QCD'' (VH-VI-231), by the  EU-Research Infrastructure Integrating Activity ``Study of Strongly
Interacting Matter" (HadronPhysics2, grant n. 227431) under the Seventh Framework Program of EU and by the
DFG (TR 16). F.H. is grateful to the COSY FFE grant No. 41788390 (COSY-058).
\end{theacknowledgments}

\bibliographystyle{aipproc} % if natbib is missing

\end{document}